\RequirePackage{fix-cm}
\documentclass[smallcondensed]{svjour3}     
\smartqed  
\usepackage{graphicx}
\begin{document}

\title{Strangeness production in antiproton-nucleus collisions}

\titlerunning{Strangeness production in antiproton-nucleus collisions}        

\author{A.B. Larionov \and
        T. Gaitanos \and
        U. Mosel
}

\authorrunning{A.B. Larionov, T. Gaitanos, U. Mosel} 

\institute{A.B. Larionov \and T. Gaitanos \and U. Mosel \at
           Institut fuer Theoretische Physik, Universitaet Giessen \\
           D-35392 Giessen, Germany \\
           \email{Alexei.Larionov@theo.physik.uni-giessen.de}
           \and
           A.B. Larionov \at
           Russian Research Center ``Kurchatov Institute'', 123182 Moscow, Russia
}


\maketitle

\begin{abstract}
Antiproton annihilations on nuclei provide a very interesting way
to study the behaviour of strange particles in the nuclear medium.
In low energy $\bar p$ annihilations, the hyperons are produced mostly
by strangeness exchange mechanisms. Thus, hyperon production
in $\bar p A$ interactions is very sensitive to
the properties of the antikaon-nucleon interaction in nuclear
medium. Within the Giessen Boltzmann-Uehling-Uhlenbeck
transport model (GiBUU), we analyse the experimental data
on $\Lambda$ and $K^0_S$ production in $\bar p A$ collisions at $p_{\rm lab}=0.2-4$
GeV/c. A satisfactory overall agreement is reached,
except for the $K^0_S$ production in $\bar p+^{20}$Ne collisions
at $p_{\rm lab}=608$ MeV/c, where we obtain substantially larger $K^0_S$ 
production rate.
We also study the $\Xi$ hyperon production, important in view of the 
forthcoming experiments at FAIR and J-PARC.
\end{abstract}

\section{Introduction}
\label{intro}
About two decades ago several experiments have been done  
on strangeness production in $\bar p$-nucleus reactions. 
In the hydrogen bubble chamber experiment at BNL \cite{Condo:1984ns},
the $\Lambda$-hyperon production in collisions 
$\bar p(0-450 {\rm MeV/c})+^{12}{\rm C},^{48}{\rm Ti},^{181}{\rm Ta}$ and $^{208}$Pb
has been measured. At KEK \cite{Miyano:1984dc,Miyano:1988mq},
the hydrogen bubble chamber measurements have been performed
for the $K^0_S, \Lambda$ and $\bar\Lambda$ production from 
$\bar p(4~{\rm GeV/c})\\ +^{181}{\rm Ta}$. Finally, at LEAR \cite{Balestra:1987vy}, 
the $K^0_S$ and $\Lambda$ rapidity yields from $\bar p(607~{\rm MeV/c})+\\ ^{20}{\rm Ne}$
interactions have been studied using the streamer chamber filled
with a natural $^{20}$Ne gas. Recently, $K^\pm$ production in annihilation 
of stopped $\bar p$'s on p, d, $^{3}$He and $^{4}$He has been measured at 
LEAR \cite{Bendiscioli:2009zza} applying the magnetic spectrometer Obelix.  

These experiments revealed some interesting features which still remain
to be explained by theory:
\begin{itemize}

\item Large ratio of the particle yields $\Lambda/K^0_S=2-3$ both for light ($^{20}$Ne)
and heavy ($^{181}$Ta) targets.

\item $\Lambda$ rapidity spectrum peaked close to the target rapidity even
for energetic collisions $\bar p(4~{\rm GeV/c})+^{181}{\rm Ta}$.

\item Enhanced strangeness production for $\bar p$ annihilations at rest involving 
more than one nucleon (baryon number of annihilating system $B>0$).

\end{itemize}
In order to describe the yields of strange particles and, in particular,
the large $\Lambda/K^0_S$ ratio in $\bar p + ^{181}{\rm Ta}$ interactions at 4 GeV/c,
Rafelski \cite{Rafelski:1988wn} assumed an annihilation fireball 
in the state of a supercooled quark-gluon plasma (QGP)
propagating through the target nucleus and absorbing nucleons.
Eventually, this fireball reaches a high baryon number, $B \simeq 10$,
and then hadronizes producing the excess of strange particles.
Such an exotic scenario has been questioned in the following-up
theoretical work by Cugnon et al. \cite{Cugnon:1990xw}, 
where the intranuclear cascade (INC) calculations of the strangeness 
production in $\bar p$-nucleus interactions have been performed.

In this talk, we present the results of our transport-theoretical 
analysis of experimental data on strangeness production from 
$\bar p$-nucleus interactions in-flight. We also make predictions on the
$\Xi$-hyperon ($S=-2$) and $\Lambda$-hypernuclear production. Section~\ref{model}    
contains a brief description of the GiBUU model applied in our calculations.
In Section~\ref{results}, we present the numerical results and discuss them. 
The summary is given in Section~\ref{summary}.

\section{Model}
\label{model}

The GiBUU transport model \cite{gibuu} solves the coupled system
of relativistic kinetic equations for the different sorts of hadrons:
\begin{equation}
  (p_0^*)^{-1}
  [ p^*_\mu \partial_x^\mu + (p_\mu^* F_i^{\nu\mu} 
                                   + m_i^*(\partial_x^\nu m_i^*)) \partial^{p*}_\nu ]
    f_i(x,{\bf p^*}) = I_i[\{f\}]~,                \label{GiBUU}
\end{equation}
where $f_i(x,{\bf p^*})$ with $x \equiv (t,{\bf r})$ denotes the distribution function of the
particles of sort $i$ ($i=N,~\bar N,~\Delta,~\bar\Delta,~Y,~\bar Y,~\Xi,~\bar\Xi,~\pi,~\eta,~\omega,~\rho,~K,~\bar K,...$
taking into account isospin projections) in the six-dimensional phase space $({\bf r},{\bf p^*})$.
If the r.h.s. of Eq.~(\ref{GiBUU}), i.e. the collision term $I_i[\{f\}]$, would be zero, 
then Eq.~(\ref{GiBUU}) would be a classical Vlasov equation describing the
propagation of the particles in the mean field potentials of a nuclear and electromagnetic 
nature.

The relativistic form of Eq.~(\ref{GiBUU}) needs to explain some specific quantities:
$p^{*\mu}=p^\mu-V_i^\mu$ is the kinetic four-momentum, $F_i^{\mu\nu} \equiv \partial^\mu V_i^\nu - \partial^\nu V_i^\mu$ 
is the field tensor, and $m_i^*=m_i+S_i$ is the effective mass. The particles are assumed
to be on their in-medium mass shells, $p^*_\mu p^{*\mu}=m_i^{*2}$. The scalar and vector
fields are expressed, respectively, as  $S_i=g_{\sigma i}\sigma$ and  
$V_i^\mu = g_{\omega i} \omega^\mu + g_{\rho i} \tau^3 \rho^{3\mu} + q_i A^\mu$. Here, $\sigma$ ($I=0,J=0$),
$\omega^\mu$ ($I=0,J=1$) and $\vec{\rho}^\mu$ ($I=1,J=1$) are the mean mesonic fields and
$A^\mu=(A^0,\bf{0})$ is the Coulomb field.

The mean mesonic and Coulomb fields are calculated, respectively, from the static 
Klein-Gordon-like and Poisson equations with the source terms provided by the space- and 
time-dependent particle densities and currents. The meson-nucleon coupling constants and the
self-interaction parameters of the $\sigma$-field are adopted from the NL3 version
of the non-linear Walecka model \cite{Lalazissis:1996rd}. The coupling constants
of the other baryons with mean mesonic fields are obtained from the corresponding 
meson-nucleon coupling constants by simple rescaling taking into account the 
light-quark contents, and, for antibaryons, the G-parity symmetry  and phenomenological
depth of the antiproton optical potential (${\rm Re}(V_{\rm opt}) \simeq -150$ MeV in the nuclear center).
This leads, e.g., to a $\Lambda$($\bar\Lambda$) potential of $\simeq -40(-450)$ MeV and $K(\bar K)$ 
potential of $\simeq -20(-220)$ MeV 
(see refs. \cite{Mishustin:2004xa,Larionov:2009tc,Gaitanos:2010fd} for further details 
of the mean fields).

The collision integral in the r.h.s. of Eq.~(\ref{GiBUU}) describes 
the antibaryon-baryon annihilation, elastic and inelastic hadron-hadron 
scattering processes, and resonance decays.
It includes the elementary hadron-hadron cross sections and resonance widths
as parameters.
The elementary cross sections are obtained by phenomenological fits to
experimental data or given by theoretical calculations.

The following {\it antibaryon-baryon} reaction channels are included:
$\bar B B \to \mbox{mesons}$ simulated by statistical annihilation model \cite{PshenichnovPhD},
$\bar B B \to \bar B B$ (elastic and charge exchange), 
$\bar N N \leftrightarrow \bar N \Delta (\bar\Delta N)$,
$\bar N N \to \bar\Lambda \Lambda$, 
$\bar N N (\bar N \Delta, \bar\Delta N) \to \bar\Lambda \Sigma (\bar\Sigma \Lambda)$,
$\bar N N (\bar N \Delta,\\ \bar\Delta N)  \to \bar\Xi \Xi$.
For invariant energies $\sqrt{s} > 2.4$ GeV ($p_{\rm lab} > 1.9$ GeV/c for
$\bar N N$), the inelastic production in antibaryon-baryon collisions,
$\bar B_1 B_2 \to \bar B_3 B_4 + \mbox{mesons}$, is simulated with the help
of the FRITIOF model.

The {\it baryon-baryon} reactions implemented in the model are: 
$B B \to B B$ (elastic and charge exchange), $NN \leftrightarrow NN\pi$, 
$NN \leftrightarrow \Delta\Delta$, $NN \leftrightarrow NR$, where
$R$ denotes any nonstrange baryonic resonance,
$N(\Delta,N^*) N(\Delta,N^*) \to N(\Delta)YK$,
$Y N \to Y N$, $\Xi N \to \Lambda \Lambda$, $\Xi N \to \Lambda \Sigma$,
$\Xi N \to \Xi N$. For $\sqrt{s} > 2.4$ GeV, the inelastic production 
in baryon-baryon collisions, $B_1 B_2 \to B_3 B_4 + \mbox{mesons}$,
is simulated by using the PYTHIA model.

The {\it meson-baryon} collisions taken into account in the model are:
$\pi N \leftrightarrow R$, $\pi N \to K \bar K N$, 
$\pi(\eta,\rho,\omega) N \to YK$, 
$\bar K N \leftrightarrow Y^*$, $\bar K N \to \bar K N$,
$\bar K N \leftrightarrow Y \pi$, $\bar K N \leftrightarrow Y^* \pi$,
$\bar K N \to \Xi K$. For $\sqrt{s} > 2.2$ GeV, the meson-baryon collisions are
simulated by applying the PYTHIA model.

The GiBUU model also includes the strangeness production/absorption
channels in meson-meson collisions: $MM \leftrightarrow \bar K(\bar K^*) K(K^*)$,
where $M$ denotes any nonstrange meson ($M=\pi,~\eta,~\eta',~\sigma,~\rho,~\omega,$).
Further details on the elementary cross sections included
in the model can be found on the GiBUU web-site \cite{gibuu}
and in the review article \cite{Buss:2011mx}.

\section{Results}
\label{results}

\begin{figure}
\begin{center}
  \includegraphics[scale = 0.50]{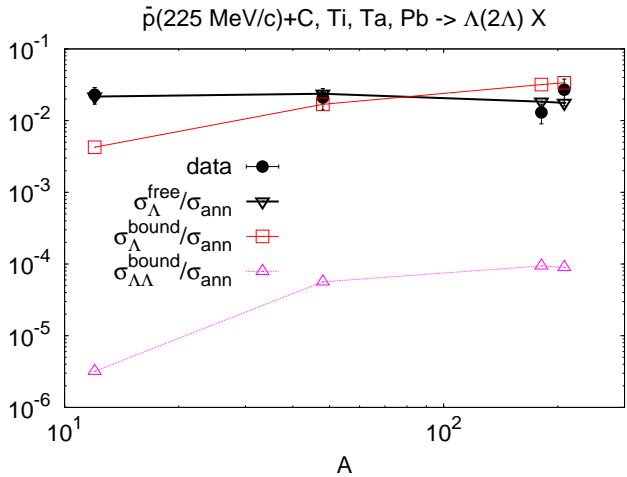}
\end{center}

\vspace*{0.5cm}

\caption{The cross sections of the free $\Lambda$ production, bound $\Lambda$ production,
and two bound $\Lambda$'s production for $\bar p (225~{\rm MeV/c})+^{12}{\rm C},~^{48}{\rm Ti},~^{181}{\rm Ta}$ and 
$^{208}$Pb. The cross sections are normalized on the annihilation cross sections of
$\bar p$ on the corresponding nuclei. The data for free $\Lambda$'s \cite{Condo:1984ns}
are for the beam momentum range 0-450 MeV/c.}
\label{fig:sig_Lambda_vs_A}
\end{figure}
We start with the low beam momenta and consider the annihilations of $\bar p$ 
at 225 MeV/c on several target nuclei.
Figure~\ref{fig:sig_Lambda_vs_A} shows the probability of the $\Lambda$ production
per $\bar p$ annihilation event on a nucleus as a function of the nucleus mass number.
We have separated the $\Lambda$ hyperons emitted to free space from those bound
in the nuclear remnant at the end of time evolution ($\simeq 200$ fm/c starting from
$\bar p$ at the longitudinal distance of nuclear radius + 5 fm from the nuclear centre).
Our results on free $\Lambda$'s agree quite well with experiment.
The probability of the bound $\Lambda$- ($\Lambda\Lambda$-) nuclear system production,
as expected, grows with the target mass number reaching $\simeq 3\%~(0.01\%)$ for the
lead target.

\begin{figure}
\begin{center}
   \includegraphics[scale = 0.50]{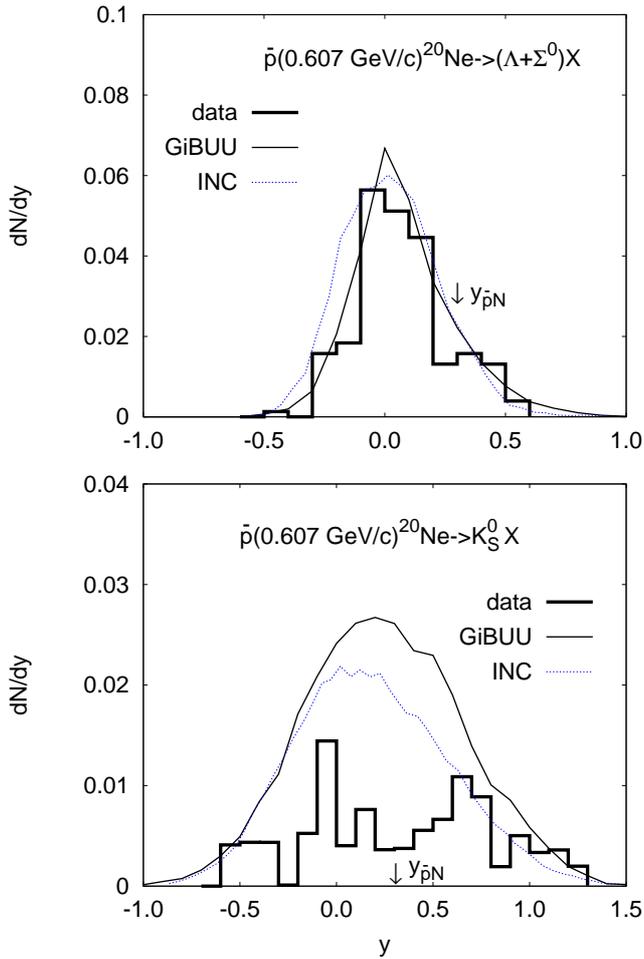}
\end{center}

\vspace*{0.5cm}

\caption{\label{fig:dsig_dy_pbarNe} The rapidity distributions of the $(\Lambda+\Sigma^0)$
hyperons and $K^0_S$ from $\bar p(607~{\rm MeV/c})+^{20}$Ne annihilations.
The distributions are normalized so that their integrals over $y$ give the number
of particles per annihilation event.
The INC calculations from \cite{Cugnon:1990xw} are also shown.
The data (histograms) are from ref. \cite{Balestra:1987vy}. Vertical arrows indicate
the rapidity of a $\bar p$-nucleon center-of-mass (c.m.) system.}
\end{figure}
In Fig.~\ref{fig:dsig_dy_pbarNe}, we present the rapidity spectra of $(\Lambda+\Sigma^0)$
and $K^0_S$ for $\bar p$ annihilations on $^{20}$Ne at 607 MeV/c. Since the $\Sigma^0 \to \Lambda \gamma$
decay is not taken into account in GiBUU, we have summed up both $\Lambda$ and $\Sigma^0$
spectra in order to compare with experimental data.
The data on $\Lambda$-rapidity yields are described quite well by GiBUU.
By studying the $Y(Y^*)$ production rate more in-detail,
we have found that $\sim 80\%$ of this rate is due to the strangeness exchange reactions 
of the type $\bar K N \to Y \pi$.

The calculated $K^0_S$ yield, $N_{K^0_S}=(N_{K^0}+N_{\bar K^0})/2$, is considerably higher 
than experiment, as one can see from Fig.~\ref{fig:dsig_dy_pbarNe}.
The INC calculations \cite{Cugnon:1990xw} are also
in a rather good agreement with $\Lambda$ rapidity spectrum, but overpredict 
the $K^0_S$ production.

\begin{figure}
\begin{center}
   \includegraphics[scale = 0.80]{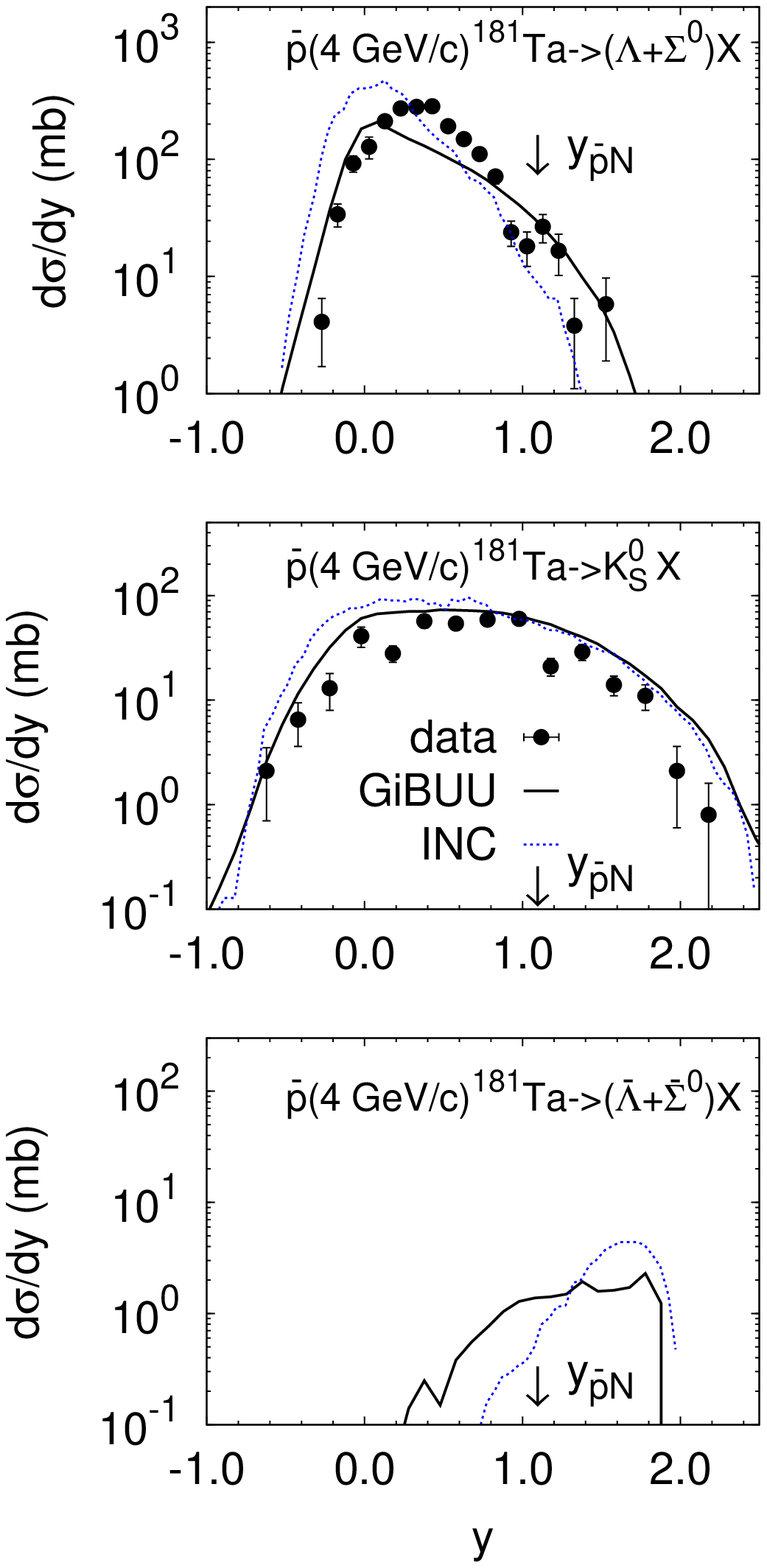}
\end{center}

\vspace*{0.5cm}

\caption{\label{fig:dsig_dy_pbarTa} 
The $(\Lambda+\Sigma^0)$, $K^0_S$ and $(\bar\Lambda+\bar\Sigma^0)$ rapidity distributions 
for the collisions $\bar p(4~{\rm GeV/c})^{181}$Ta. The INC calculations
are taken from \cite{Cugnon:1990xw}. The data are from 
\cite{Miyano:1988mq}.}
\end{figure} 
Fig.~\ref{fig:dsig_dy_pbarTa} presents the $(\Lambda+\Sigma^0)$, $K^0_S$
and $(\bar\Lambda+\bar\Sigma^0)$ rapidity spectra from $\bar p+^{181}$Ta collisions at 4 GeV/c. 
The GiBUU model reproduces the tails of the $y$-distribution of $\Lambda$'s 
fairly well, but underestimates the data at the peak position, $y \simeq 0.3$.
On the other hand, the INC model overestimates the data around the target 
rapidity, $y=0$. Both models produce the peak position of the $\Lambda$ rapidity 
spectrum at $y \simeq 0$. The theoretical rapidity spectra of $K^0_S$
agree with data reasonably well, except for target rapidities,
where some excess of $K^0_S$ is still visible.
By inspecting the $Y(Y^*)$ production rate we again found that $\sim 70-80\%$
of this rate is due to the strangeness exchange processes 
$\bar K(\bar K^*) B \to Y X$, $\bar K B  \to Y^*$, and $\bar K B  \to  Y^* \pi$, while the
total contribution of the $\bar B B$ collision channels --- including
the direct $\bar N N$ channel --- is on the level of a few percent only.
The difference between the $(\bar\Lambda+\bar\Sigma^0)$ rapidity spectra 
produced by GiBUU and INC is mostly due to somewhat different
angular distributions for the $\bar N N \to \bar Y Y$ processes and due
to different $\bar Y N$ annihilation cross sections used in the both models.
We also note that the INC model does not contain mean field potentials,
while our calculations include strongly attractive antihyperon potentials. 

%
%

Let us, finally, discuss the predictions of the GiBUU model for the $S=-2$ hyperon
production in $\bar p+^{197}$Au collisions at 3 GeV/c. We have found that the predominant
channel of $\Xi$-hyperon production is the collisions of the strange mesons $\bar K,~\bar K^*,~K$ 
and $K^*$ with baryons which are responsible for $\sim 35 \%$ of the total $\Xi$ yield. 
Almost of the same importance are the $\Xi^* \to \Xi \pi$ decays ($\sim 26 \%$), and strange meson%
-hyperon collisions ($\sim 17 \%$).
The direct channel $\bar N N \to \bar\Xi \Xi$ is responsible only for a quite small fraction
of produced $\Xi$'s ($\sim 5 \%$).

\begin{figure}
\begin{center}
   \includegraphics[scale = 0.80]{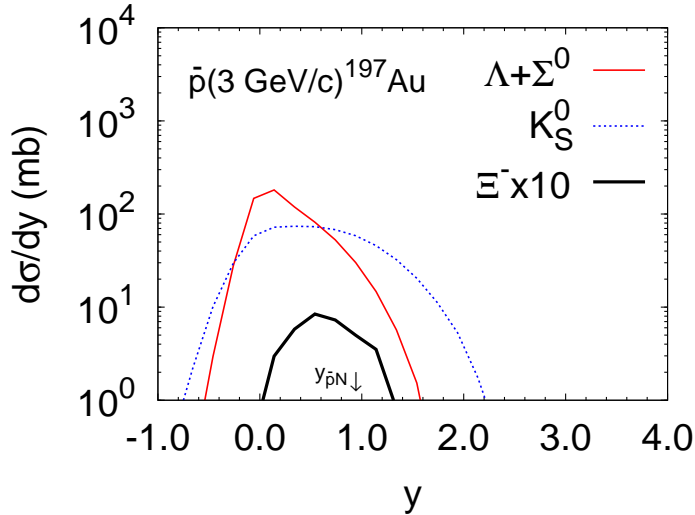}
\end{center}

\vspace*{0.5cm}

\caption{\label{fig:dsig_dy_pbarAu_3gevc} Rapidity distributions of $(\Lambda+\Sigma^0)$,
$\Xi^-$ and $K^0_S$ from $\bar p$ interactions with $^{197}$Au at 3 GeV/c. The spectrum
of $\Xi^-$ hyperons is multiplied by a factor of 10.}
\end{figure}
Fig.~\ref{fig:dsig_dy_pbarAu_3gevc} shows the rapidity spectra of 
$\Xi^-$, $(\Lambda+\Sigma^0)$ and $K^0_S$ for the $\bar p(3~{\rm GeV/c})+\\ ^{197}{\rm Au}$ collisions. 
The spectra of the $(\Lambda+\Sigma^0)$ hyperons are peaked close to $y=0$. 
The $K^0_S$ spectra have also a maximum at $y \simeq 0$ spread
towards forward rapidities. This behaviour has been interpreted
in \cite{Rafelski:1988wn} as a signature of strangeness production from a QGP fireball.
In this picture, the rapidity spectra of all strange particles emitted from 
the fireball should have their maxima at the c.m. rapidity of the fireball. 
However, the spectrum of $\Xi^-$ hyperons produced by GiBUU calculations is shifted
to the forward rapidities and peaked at $y \simeq 0.6$. The difference between the $(\Lambda+\Sigma^0)$
and $\Xi^-$ rapidity spectra can be understood in our pure hadronic transport calculations 
as the consequence of different thresholds for $Y$ and $\Xi$ production in 
antikaon-nucleon collisions. Indeed, the $S=-1$ hyperons are mostly produced in exothermic 
reactions, like $\bar K N \to Y \pi$, with slow initial $\bar K$. As a result, the produced hyperon 
$Y$ is also slow and its momentum is isotropically distributed in the laboratory frame. 
However, the double strangeness exchange process, $\bar K N \to \Xi K$,
mainly responsible for $\Xi$ production, is endothermic ($p_{\rm lab}^{\rm thr}=1.05$ GeV/c --- $\bar K$ 
beam momentum at threshold, $y_{c.m.}^{\rm thr}=0.55$ --- c.m. rapidity at threshold) and requires
a fast incoming $\bar K$. Such antikaons are mostly emitted in the forward direction in the
laboratory frame. Thus, the outgoing $\Xi^-$ moves also forward due to the c.m. motion of the
$\bar K N$ system.

\section{Summary}
\label{summary}

We have performed microscopic transport calculations of strangeness production
in antiproton-nucleus collisions at $p_{\rm lab}=0.2-4$ GeV/c on the basis of the 
GiBUU model. The main results can be summarized as follows:
\begin{itemize}

\item The experimental data on $\Lambda$-yields are reasonably well described.
However, there are some deviations in detailed shape of $\Lambda$-rapidity 
spectrum at 4 GeV/c. The $K^0_S$-yields are overestimated. The yield ratios 
for $\bar p+^{20}$Ne at 608 MeV/c are: $\Lambda/K^0_S=1.0$ (GiBUU),  $\Lambda/K^0_S=1.4$ (INC),
$\Lambda/K^0_S=2.3 \pm 0.7$ (exp.). This indicates missing antikaon absorption 
by $\bar K N \to Y \pi$ in transport calculations.

\item The peak positions of the $(\Lambda+\Sigma^0)$- and $\Xi$-hyperon rapidity
spectra strongly differ. In future experiments at FAIR and J-PARC, 
this can be tested in order to support or exclude the possible exotic mechanism 
of strangeness production via a QGP fireball.

\item $\Lambda$- and $\Lambda\Lambda$-hypernuclei production is possible on the primary 
target at low $\bar p$-beam momenta.

\end{itemize}

\begin{acknowledgements}
The support by the Frankfurt Center for Scientific Computing is gratefully 
aknowledged.
This work was financially supported by the Bundesministerium f\"ur Bildung und Forschung,
by the Helmholtz International Center for FAIR within the framework of the LOEWE program,
and by the Grant NSH-7235.2010.2 (Russia).
\end{acknowledgements}




\bibliographystyle{spphys}       

\bibliography{larionov_leap2011}   

\end{document}